# INCIDENT RESPONSE PLAN FOR A SMALL TO MEDIUM SIZED HOSPITAL

Charles DeVoe[1] and Syed (Shawon) M. Rahman, Ph.D.[2]

[1] Capella University, Minneapolis, USA
cdevoejr@capellauniversity.edu

[2]Computer Science Faculty, University of Hawaii-Hilo, Hilo, USA
and Adjunct Faculty, Capella University, Minneapolis, USA

SRahman@Hawaii.edu

## ABSTRACT

*Most small to medium health care organizations do not have the capability to address cyber incidents within the organization. Those that do are poorly trained and ill equipped. These health care organizations are subject to various laws that address privacy concerns, proper handling of financial information, and Personally Identifiable Information. Currently an IT staff handles responses to these incidents in an Ad Hoc manner. A properly trained, staffed, and equipped Cyber Incident Response Team is needed to quickly respond to these incidents to minimize data loss, and provide forensic data for the purpose of notification, disciplinary action, legal action, and to remove the risk vector. This paper[1] will use the proven Incident Command System model used in emergency services to show any sized agency can have an adequate CIRT.*

## 1 Introduction

Many small to medium health care organizations often fail to see the need for a Computer Incident Response Team (CIRT) or feel they do not have the resources needed to implement one. This paper will address the need and justification for a CIRT as well as explain how to implement one with limited resources. First there will be a discussion considering the legal implications which also addresses the ethical needs for protecting patient information including Personally Identifiable Information (PII). Subsequent to this a brief overview of the Incident Response process will be given. The following sections will describe the methodology, the Emergency Services Incident Command System (ICS), followed by an adaption of the ICS to the CIRT. The paper ends with a discussion of the skill set required by CIRT members and the tools they will need.

---

[1] This work is partially supported by EPSCoR award EPS-0903833 from the National Science Foundation (NSF) to the University of Hawaii, USA

It is important to recognize that no institution is exempt from attack or compromise. In January of 2012 the Multi-State Information Sharing and Analysis Center (MS-ISAC) published its list of emerging trends and threats for 2012 [ HYPERLINK \l "Pel12" **1** ]. The MS-ISAC lists the following emerging trends and threats:

- Mobile devices and Applications – The use of mobile devices and applications continue to grow. As they grow so do the attacks and the attack vectors.
- Hacktivism – Attacks carried out by activist to send a political or social message. Groups include Anonymous and Lulz.
- Search Engine Optimization (SEO) Poisoning - Cyber criminals will take advantage of the 24-hour news cycle to target visitors searching on the most popular keywords or sites and infect users via sites designed to look like legitimate news services, Twitter feeds, Facebook posts/emails, LinkedIn updates, YouTube video comments, and forum conversations. We expect cyber criminals to take advantages of notable news events such as the London Olympics, U.S. presidential elections, and Mayan calendar predictions.
- Social Engineering - The use of rogue anti-virus to entice users into clicking on malicious links, fake registry cleanup, fake speed improvement software, and fake back-up software mimicking popular personal cloud services.
- Advanced Persistent Threat - a long-term pattern of targeted hacking attacks using subversive and stealthy means to gain continual, persistent exfiltration of intellectual capital.
- Spear Phishing Attacks – These attacks are targeted at particular individual and appear to be legitimate emails. This attacks are typically conducted by a person seeking financial gain, trade secrets or sensitive information

Each and every one of these threats has a human element to it. Humans unlike machines, do make mistakes, are prone to errors, and can be coerced into giving out information that is personal in nature or proprietary. Given all of this, there is no reason to believe that the number of incidents will decline; in fact, given the ever increasing sophistication of the hackers it is believed the number of incidents will increase **1**].

In April of 2011 it was reported that since 2009, 10.8 million people have been affected by data breaches according to The U.S. Department of Health and Human Services. These breaches occurred in 265 separate incidents [ HYPERLINK \l "How11" **2** ]. Incidents include things like loss of equipment, improperly handling obsolete data, attack from a hacker, phishing, social engineering, and infection via email or visiting a malicious web site. Once an incident occurs it is critical that it be handled properly to determine the extent of the problem and to minimize the impact.

The Information Technology (IT) group will typically focus on making the system operational again. That is, they will rebuild the system, clean the system to remove the infection, restore the system from backup, block the attacking vector, or a combination of these. There is much more that is needed to properly handle, the incident and ensuing investigation; forensic data is required to determine what data if any was taken, by whom, and how. Obtaining the evidence allows the organization to determine what entities have been affected. Additionally, it addresses if there is the need for personnel actions, if law enforcement should be involved, or both. This information can then be used by the organization to patch holes; change polices; or fix the problem.

For these reasons a dedicated Cyber Incident Response Team (CIRT) is needed to properly respond to incidents and gather evidence. A CIRT focuses on six functions Preparation, Identification, Containment, Eradication, Recovery, and Follow Up **3**]. These functions will assure a systematic and appropriate response enabling a rapid and thorough recovery, minimizing the loss of data and downtime. Following proper procedures will preserve system data such that a forensic analysis can be done. This analysis will provide details of how, when, where, and what. Knowing this the organization can prevent future incidents over the same attack vector and detect evidence to support personnel and legal actions. This data is also necessary to support reporting requirements to individuals as well as regulatory agencies. It is imperative that senior management supports the effort and provides proper authority to operate the CIRT.

## 1.1   Background of the Study

Most health care organizations have implemented and installed preventive measures that include firewalls at the border, Intrusion Protection System (IPS), host based Intrusion Detection System (IDS), anti-virus software, WEB content filtering (i.e. WebSense), and switched networks (segmenting functional groups). End users have been trained in proper procedure and the safeguarding of information. Strong, effective, policies have been developed to protect information. Even with all of this protection, breaches, infections, and human errors will still happen. Equipment is lost, stolen, or misplaced. Anti-virus software only detects 50% of all viruses and is vendor dependent [  HYPERLINK \l "Vir12"  **4** ]. New viruses are constantly being developed and old ones are modified to avoid detection. Not only must the hospital protect corporate data and systems, it must also protect industrial control systems, and the physical access electronic. Even the best protection plans can be breached.

When an incident is discovered in the current system an ad hoc response is executed. System administrators and technicians who are untrained in incident response will first attempt to get the system operational before attempting to preserve evidence or determine the extent of the breach.

## 2   Legal and Ethical Implications

While many laws apply to computer crimes special attention must be given to the rights and protections of the individual when responding to an incident. Of primary concern here are the 4th amendment protections of the U.S. Constitution, the Computer Fraud and Abuse Act, and the Electronic Communications Privacy Act. Each incident should be approached form the point of view that a crime has been committed; this provides the highest level of diligence and assures proper evidence gathering. It is imperative that the rights of the individual be protected so as to not violate due process.

Additionally, the responder must be aware of their responsibilities and limits; one could actually face criminal or civil penalties or legal action if proper procedures are not followed. This section will address these concerns. The content in this section is not meant to be an exhaustive examination of the legal issues, it is meant to bring awareness about the laws involved.

## 2.1   HIPAA

The hospital is a covered entity under the HIPAA privacy rule. As such the Protected Health Information (PHI) of patients must be safeguarded and handled appropriately. Improperly handing this information creates an incident. PHI is any information that relates to the person's past, present and future health condition, the care received, or anything to do with payments. Employment records are exempt **5**]. There are no restrictions on de-identified information.

In general the data may be used and released to the individual; for treatment, payment, and health care operations; for the public interest and benefit; and limited data sets where the PHI has been removed. In the case of the public interest or benefit, PHI may be released to law enforcement; workers compensation; when there is a serious threat to health and safety; Health oversight activities; Victims of abuse neglect or domestic violence; and for Public Health Activities [ HYPERLINK \l "HIP09" **6** ]. PHI is not to be used for marketing or any other unauthorized purpose.

## 2.2   HITECH Act

The HITECH Act was implemented as part of the American Recovery and Reinvestment Act of 2009. Under this act HIPAA was strengthened to include fines, and a data breach notification **7**]. In order to determine if a breach actually occurred the Hospital must perform an investigation to determine what data may have been breached. Civil Penalties are listed in Table 1:

Table 1 – List of HITECH Civil Penalties [  HYPERLINK \l "HIP09"  **6**  ]

| Violation category | Each violation | All such violations of an identical provision in a calendar year |
|---|---|---|
| **Did Not Know** | $100–$50,000 | $1,500,000 |
| **Reasonable Cause** | $1,000–$50,000 | $1,500,000 |
| **Willful Neglect—Corrected** | $10,000–$50,000 | $1,500,000 |
| **Willful Neglect—Not Corrected** | $50,000 | $1,500,000 |

## 2.3   Amendment IV

The right of the people to be secure in their persons, houses, papers, and effects, against unreasonable searches and seizures, shall not be violated, and no warrants shall issue, but upon probable cause, supported by oath or affirmation, and particularly describing the place to be searched, and the persons or things to be seized.

The fourth amendment protections only apply in cases where law enforcement or the government are involved **8**]. Government involvement includes actions performed on behalf of a government representative. As an example of this, a system administrator can view the files on a server of a user and there are no fourth amendment protections. However, if a member of law enforcement asks the same administrator to view those files the administrator is now working at the direction of law enforcement and fourth amendment protections apply. In this case a search warrant is required.

It should also be noted that there are no Fourth Amendment protections in all cases where the government is involved. For there to be a fourth amendment issue there must first be a reasonable expectation of privacy. In Katz v. United States, 389 U.S. 347 (1967), the Supreme Court held that the Fourth amendment protects people not places. Justice Stewart issuing the majority opinion stated: "What a person knowingly exposes to the public, even in his own home or office, is not a subject of Fourth Amendment protection. But what he seeks to preserve as private, even in an area accessible to the public, may be constitutionally protected." [ HYPERLINK \l "Joh67" **9** ] There are also no fourth amendment protections for items that are in plain view.

In the age of high technology with enhanced viewing capabilities the question of using thermal imaging devices, highly sensitive listening systems, Satellite imagery, etc. have come into question. In KYLLO V. UNITED STATES **10**] Kyllo held that the use of a thermal imaging camera to detect heat emanating from the home was a violation of the fourth amendment. In this case Agent William Elliott used an infrared camera from the public street to scan the residence to detect the use if high intensity lights in an indoor marijuana growing operation. The government argued that the camera only measured the IR coming from the part of the structure that was in plain view and thus did not violate any rights. However, the justices applied Katz to this ruling stating, "We rejected such a mechanical interpretation of the Fourth Amendment in Katz, where the eavesdropping device picked up only sound waves that reached the exterior of the phone booth. " In concluding Justice Scalia stated "Where, as here, the Government uses a device that is not in general public use, to explore details of the home that would previously have been unknowable without physical intrusion, the surveillance is a "search" and is presumptively unreasonable without a warrant."

Interestingly in DOW CHEMICAL CO. v. UNITED STATES [ HYPERLINK \l "Bur86" **11** ] the courts found that the use of aerial photography was acceptable: The Court notes that EPA did not use "some unique sensory device that, for example, could penetrate the walls of buildings and record conversations." Nor did EPA use "satellite technology" or another type of "equipment not generally available to the public."

For the incident responder at this company fourth amendment rights are not of much concern unless law enforcement or the government are involved. In those cases it the responder needs to be aware of these restrictions and assure that law enforcement is acting appropriately. It is imperative that the employees react appropriately to request from law enforcement; should there be questions or issues the request should be forwarded to the legal department and senior management.

## 2.4    Computer Fraud and Abuse Act

This is the first law specifically targeting computers and computer systems. The law targets systems under direct control of a federal entity, financial institution, or computers involved in commerce. This law made it a crime to gain unauthorized access to computer system to gain information concerning national security, financial records, or information from any agency within United States government. Originally passed in 1984 this law has seen several revisions.

The Computer Fraud and Abuse Act (CFAA) covers crimes where the perpetrator has "knowingly accessed a computer without authorization or exceeding authorized access," 12]. It is generally accepted that without authorization pertains to outsiders (hackers) while exceeding authorized access applies to insiders. The CFAA covers seven areas:

- obtaining national security information
- compromising confidentiality
- trespassing in a government computer
- accessing to defraud and obtain value
- damaging a computer or information
- trafficking in passwords
- threatening to damage a computer

The incident responder needs to be highly cognizant and fully aware if this law for two reasons. The responder must assure that they do not break the law by exceeding their authority or authorized access. Secondly, the responder must be able to identify potential acts of criminal activity.

## 2.5   Electronic Communications Privacy Act

### 2.5.1   THE ELECTRONIC COMMUNICATIONS PRIVACY ACT

This law deals with the "Interception and disclosure of wire, oral, or electronic communications" as stated in 18 U.S.C. § 2510-22 (Title I) (United States Code: Title 18,CHAPTER 119); Title I of this law prohibits the interception of electronic communications. It also prohibits the use and manufacture of devices intended to intercept electronic communications. However as with most laws there are exceptions; the law provides Internet Service Providers (ISP) with the authority to intercept communications "in the normal course of his employment while engaged in any activity which is a necessary incident to the rendition of his service". This law also provides procedures for law enforcement in the event that surveillance of electronic communications is required. It also prohibits the use of illegally obtained information as evidence. This law was originally passed in 1986.

### 2.5.2   STORED COMMUNICATIONS ACT

18 U.S.C. § 2701-12 (Title II) covers "Stored wire and electronic communications and transactional records access" (United States Code: Title 18, CHAPTER 121). This law addresses unauthorized access (including privilege escalation) for users who "…obtains, alters, or prevents authorized access to a wire or electronic communication while it is in electronic storage". This law also has exceptions allowing the owner of the system for maintenance, backups, and other needed services. Of interest in this section is that if the owner the system finds evidence of "…an emergency involving danger of death or serious physical injury to any person" or in cases of child exploitation. This law was originally passed in 1986.

## 2.6   Security Breach  Notification

According to The National Council of State Legislatures "Forty-six states, the District of Columbia, Guam, Puerto Rico and the Virgin Islands have enacted legislation requiring notification of security breaches involving personal information." [ HYPERLINK \l "Nat12" **13** ]. The importance here is that the organization must determine what data was breached and whom notification must go to. It is in the best interest of the organization to keep the number of notifications minimal to reduce cost and maintain a good reputation.

# 3   Cyber Incident Defined

NIST Special Publication 800-61 states, "A computer security incident is a violation or imminent threat of violation of computer security policies, acceptable use policies, or standard computer security practices."   For the purposes of this plan an incident will also be defined as any illegal activity involving Hospital computers, data or both.   This definition assumes that adequate polices are in place to protect the intellectual property of the organization; the PII of employees, customers, and partners; and Protected Health Information (PHI) of patients.

Incidents are created when events are triggered that give an indication of an adverse action. Events are things like IDS signature triggers, anti-virus detection, firewall detecting a host going to a known malicious domain, a phishing email is opened, or a firewall blocking a connection attempt.   Even something like a "slow computer" is an event.   Not all events produce incidents some are simply false positives.   An example would be a web filter blocking certain content on a web page.   For example, a user visits a legitimate site like cnn.com; on that page are various advertisements and links that redirect the user to other web sites.   If the browser attempts to pre-fetch the links and resolve them it may appear an end user attempted to access an unauthorized web site.   In reality, the user did no such thing.

Incidents can occur at random times and there can be multiple incidents occurring at the same time.   The complexity of the incident can also vary as can the duration.   A simple virus infection may only require cleaning the host while a data breach or discovery of illegal activity may take an extensive investigation and lengthy remediation.   Finally, it is entirely possible to have multiple incidents that are all connected and interrelated.   As can be seen, the response to the incident may vary in size, scope, and necessary resources.   The incident response team needs the ability to maintain control and grow as the incident grows.

In addition to the typical cyber incident (i.e. hacking, malware, data loss) there are also the response to physical attacks.   These include things like theft, fire, flood, building collapse, bomb threats, etc.

# 4   Incident Response Procedure

When responding to an incident it is important to only tell people who have a need to know about the incident. It is possible that an insider is violating law or policy.   Allowing them to know they are being watched or investigated gives them the opportunity to destroy valuable evidence. Throughout the process take good notes.

## 4.1   Preparation

In the preparation phase user expectations are set, as well as assuring system administrators are properly prepared.   This is also the place where polices are set for the IRT to include the scope, responsibilities, when to call executives management and when to call law enforcement. This plan in fact is part of the preparation phase.

As part of the preparation it is imperative that a logon warning banner be placed on all systems. This banner not only sets the expectation that end users will be monitored but it also serves notice that there is no expectation of privacy. By so doing, it makes the task of analyzing events much easier as it removes the need to get approvals from senior management.

Also addressed here is the need to notify law enforcement and when. There are times when Law enforcement must be notified, for example when there is a serious safety or life threat to the general public, in cases of child pornography, in cases of potential abuse to name a few. There are also reasons not to involve law enforcement. In so doing the organization loses control of the case; in fact there are now two cases with different objectives. The organization wishes to get systems back on line and operational while law enforcement wishes to gather evidence and prosecute a crime. In this case, the hospital also risks equipment seizure by the authorities. Finally, it may be desirable not to report the issue to eliminate negative publicity.
Policy must also be developed for critical systems such that there are regular backups. This will allow for restoration of the system should a compromise occur.

## 4.2    Identification

The goal here is to examine the events, analyze them, and determine if there is an incident. As mentioned previously, not all events are incidents. Examples of this are phishing emails that are not opened, a user on a Linux system surfing to a web site with a known windows exploit, and a large increase in ftp traffic that is authorized.

## 4.3    Containment

The purpose of this phase is to eliminate further damage. There is the short term goal and the long term goal. In the short term the desire is to stop communications with the hacker, botnet, or other external entity. Additionally, the system needs to be isolated to stop the spread of the virus or malware. Some methods of doing this are to unplug the network cable, disable the switch port, put in ACLs on routers or firewalls, and as a last resort unplug the power. Unplugging the system is not advised as this results in the loss of critical data and evidence concerning the running state of the system. Most importantly, try to keep a low profile, do not let the attacker now that you have discovered their activity.

Also at this time make a system image to include the file system, and memory. When performing forensic analysis later always use a copy of these images. Keep the original pristine for evidence purposes. Take pictures of the area and the current state of the screen.

Long term containment involves applying patches to the affected system and other similar systems. Consider changing passwords and adding firewall rules. Remove any accounts that were used by the attacker and shutdown hacker processes. Subsequent to these actions the virus or malware must be eradicated.

## 4.4    Eradication

During eradication the system is cleaned and attacker artifacts are removed. This is where the forensic analysis will help. If the malware can be identified a simple search on the internet will reveal the characteristics of the malware. It may be necessary to use a VM or sand box to run the malware and observe what it is doing.

Find the most recent clean backup to restore the system. This backup should also be checked for the presence of the malware or attacker. If a rootkit was installed this will modify the operating system itself, in this case reformat the hard drive and reinstall the operating system. Once the system is restored perform a vulnerability scan using Nessus and patch all vulnerabilities.

## 4.5    Recovery

Prior to putting the system back into production check the operation of the system against the test plan and baseline documentation. This should be done by the system administrator and the owner of the system. The owner of the system makes the final decision on putting the system back into production. Once in production monitor the system closely and check carefully for signs of re-compromise.

## 4.6    Follow Up

In the follow up phase the lessons learned are documented. A report is generated detailing how the attacker got in, what was done, how the issue was found, and what was done to fix it. It should also include recommendations to prevent future attacks by the same method.

## 4.7    Seven Deadly Sins

1. Failure to report or ask for help
2. Incomplete or nonexistent notes
3. Mishandling or destroying Evidence
4. Failure to create working images
5. Failure to contain or eradicate
6. Failure to prevent re-infection
7. Failure to apply lessons learned.

# 5    Methodology

Responding to cyber incidents is much like responding to a fire or medical emergency. In all cases the first responders must assess the situation, determine what resources may be needed, evaluate the severity of the situation, determine a strategy and tactics, and finally assure safe operations. The fire and emergency services have been responding to emergencies for many decades and have developed very good procedures and methods for doing so. The fire service uses an Incident Command System (ICS), which is proven and can be easily adapted for use in cyber incident response. The ICS is designed to handle incidents of all sizes and complexity and can grow and shrink as the incident grows and shrinks.

## 5.1    The Incident Command System (ICS)

The fire service has a well-designed and proven system that is used for responding to various incidents. Cyber incidents can be thought of as fires within computer systems and networks. As such, it is a simple task to adapt the ICS to cyber incidents.

The ICS was developed in the 1970's in response to several catastrophic fires in the urban interface in California **14**]. This system is designed to allow inter-agency operation for incidents small to large and complex. The ICS is based on following 14 management characteristics that provide strength and efficiency to the total system [ HYPERLINK \l "FEM12" **15** ].

- Common Terminology – Assures that all participants are using the proper terminology for incident response and required resources.
- Modular Organization – This allows for a compartmentalized approach that allows resources or functions to be brought in based on the complexity and severity of the incident.
- Management by Objectives – Establishes all-encompassing objectives and goals
- Incident Action Planning – Provides a centralized approach to the planning of the response to the incident as well as setting priorities.
- Manageable Span of Control - Span of control defines how many people (or things) each individual can manage. The typical range is 3 to 7 with 5 being optimal.
- Incident Facilities and Locations – Numerous and varied facilities may be needed for incident, these include command posts, staging areas, rest areas, etc.
- Comprehensive Resource Management – This means to maintain a comprehensive view of resource utilization. Resources in this case include equipment and personnel.
- Integrated Communications – Establishes a common communication system to address the equipment, systems, and protocols necessary to achieve integrated voice and data communications.
- Establishment and Transfer of Command – In the beginning of any incident command (who is n charge?) must be established. As the incident grows it may be necessary to transfer the command, this requires a briefing that includes the current status, the plan, and other important information
- Chain of Command and Unity of Command – The chain of command assures that subordinates report to supervisors. The concept actually comes form the military where it is not desirable to have privates reporting to generals.
- Unified Command – This is a concept that allows various agencies and entities with different functional, geographical and legal authority to work together.
- Accountability – This is management of personnel and resources involved in the incident.
- Dispatch/Deployment – Personnel and resources only respond when requested.
- Information and Intelligence Management – This is the process of gathering, analyzing, assessing, sharing, and managing incident-related information and intelligence.

The incident command sections are defined as follows **15**]:

- Command – Incident Commander (IC), Public Information Officer, Liaison Officer

- Operations – Manages the tactical operations
- Planning – Resources, Situation, Documentation (Understanding the situation, establish Priorities, and Strategy)
- Logistics – Communication, Food, Supply, Facilities
- Finance/Administration – Procurement, Compensations, Claims, Cost
- Intelligence/Investigations – Post incident investigation or intelligence gathering.

Not all components of the ICS are invoked at every emergency incident. Each component is invoked as needed. It is possible that the incident commander will also be the operations chief, planning chief, logistics chief, and finance chief. To show how this works the following scenarios are provided.

Scenario 1 – The incident is a small shed (10 ' x 10') on fire with no other structures, vehicles, or combustibles nearby. A 5-man team with one truck arrives; the ranking officer (Lieutenant) assesses the scene. This officer assumes the role of incident commander (Command) and develops a plan for fighting the fire (Planning). He then directs the subordinates on the tactical operations for extinguishing the fire (Operations).

Scenario 2 – A large two family home in a remote location is on fire. The first arriving officer (Lieutenant) takes command of the incident and establishes a command post. He notes the structure is 50% involved and that additional resources will be required. The incident commander (IC) calls for additional teams to be deployed. As the teams arrive a higher-ranking officer (Captain) takes over the command. In so doing he is briefed by the lieutenant before assuming command. The lieutenant is then reassigned to the operations section where he will lead an attack team. Other teams are assembled to supply water, and provide support for the operations. As this is a lengthy incident, the logistics section is established to provide refreshments and food.

No two emergencies are alike; it is possible that the emergency incident may also require law enforcement, hazardous materials or environmental experts, medical, heavy construction, or any other numerous resources. It should be noted that even if a chief is called in by captain or lieutenant the chief does not have to take over control of the incident. The chief can act in an advisory role. As can be seen, this model allows for the expansion and inclusion of these resources as needed. Additionally, recognizing the potential need for these resources allows for pre-planning and establishing contacts.

## 5.2    Converting the ICS to Cyber Space

To convert the ICS from the fire service to the cyber world is a relatively straightforward process. Of the 14 management principles 3 are modified slightly

- Incident Facilities and Locations – When responding to cyber incidents the need for a well-known command post, staging areas, rest areas, etc. are not needed. However, there is a need for secured workspaces, forensics labs, and other such facilities. Additionally, it is critical that a primary incident responder be assigned.
- Integrated Communications – A communication system may or may not be required depending on the type of incident and the sensitivity of the data. The plan should however address the need for multiple communication mediums. That is, if the system under attack is the phone systems or the email system there needs to be another means of communication available. This is also dependent upon the size of the geographic area covered.

- Dispatch/Deployment – It may be necessary to send incident responders to outreach locations or remote sites. In this case there must be control of who is sent where, with what resources and mission assignments.

As with the management functions the sections are very straightforward as well. In this section the various roles will be spelled out as they pertain to cyber incident response.

Command – The Incident Commander (IC) is responsible for managing the incident from start to finish. Transfer of authority for a multi-day incident is not required. The individual responsible for the response will be available 24 X 7, however they are not required to be on site at all times. The Public Information Officer will be utilized only on breaches that involve outside agencies or the public. The Liaison Officer is typically not utilized. The function of this officer is to facilitate communications between agencies and leverage outside resources. The liaison officer may be needed in cases where law enforcement is required or coordination between partner organizations is required.

- Operations – Manages the tactical operations
- Planning – Resources, Situation, Documentation (Understanding the situation, establish Priorities, and Strategy)
- Logistics – Communication, Food, Supply, Facilities
- Finance/Administration – Procurement, Compensations, Claims, Cost
- Intelligence/Investigations – Post incident investigation or intelligence gathering.
- 

To consider how this works in the realm of cyber incidents examine the following scenarios.

Scenario 1: The corporate anti-virus console has detected a piece of malware on an end users computer. The system administrator notifies the IRT who sends a member of the team out to investigate. The responder finds the malware, examines the system for compromise, and determines if further action is required. He reports back to the manger who concurs. The system is then cleaned by the IT department and reinstalled.

Scenario 2: While reviewing log files the network administrator notices that one workstation is downloading a large amount of data to an external IP address. The incident response team gets the log files and begins analyzing them. They find that the files are being sent to a personal Roadrunner account. This could now be an indication of a policy violation or illegal activity. To assure the investigation is handled properly the legal department is activated along with HR department. The CEO is informed of the downloads and based on the size they could potentially contain sensitive information. As the team continues to dig the find that one of the people in the engineering department is downloading facility information so that he can work on this at home. While not illegal, this is a violation of policy and will be handled by the supervisor and HR.

Scenario 3: This scenario is the same as the previous only this time the employee is a clerk who is downloading patient PPI. The PPI includes names, Social Security numbers, addresses, billing information (to include credit card numbers) and treatments. In this case not only has corporate policy been violated but laws may also have been broken. Addressing this incident requires that Legal, HR, Finance, facility security, and senior management be involved. Senior managers may decide to call in law enforcement to assure that data and evidence are collected and gathered in accordance with proper procedures. As the investigation progresses it is determined that the employee is taking the data and selling to identify thieves. This now constitutes a data breach, which requires the Public Information officer involvement. In a coordinated effort the employee will be arrested, accounts will be disabled, facility access removed, and the employee will be fired. Subsequently, press releases will be written and affected end users will be notified of the breach by the Public Information office. In this incident the CEO will be called in and will take over control. In so doing the ISO well have a meeting with the CEO who will be briefed on the status and the plan. The ISO will then focus on remediation and forensics while the CEO directs others in preparing the legal, HR, and Public Information responses.

Using the above three scenarios one can easily see how the incident response can grow and be managed in an orderly fashion. It should be noted that even though the CEO is called into an incident the CEO does not have to take over control of the incident. His function could be an advisory one. In the case of all serious incidents the CEO should be notified. One should also consider the situation where there are multiple incidents occurring at the same time. Using this model a unified command post is established with individual teams reporting back. Again, the system can grow as needed.

# 6   Incident Response Team

## 6.1   IRT Member Skills

Incident Responders should hold at least one reputable forensics certification available from organizations like SANS, GIAC Certified Forensics Analyst (GCFA), or International Association of Computer Investigative Specialist (IACIS) Certified Forensic Computer Examiner (CFCE). In the opinion of the author, the independent certifications are better than the ones offered up by vendors like EnCase.

## 6.2   Lab

When building an incident response laboratory one needs to assure that the facility is secure and only used by authorized individuals. All investigations should be treated as if they will be used for the enforcement of a crime. The key is to remember the need to know rule, thus restricting investigators to knowledge of only their own investigations. Some things to consider are:

- Locking evidence cabinets.
- Locking room
- TEMPEST Certified – Not necessary, but nice to have. In this regard it may be advantageous to implement some of the features to make the investigative room more secure.
- Workstations with ample work space (approximately 150 sq. ft.)
- Limited view of workstation screen.
- Budgeting for additional systems each year to maintain compatibility with hardware and software advancements
- An inventory of Operating Systems to include old and obsolete systems.
- Regulated Access.

There should be adequate numbers of forensics workstations. It is recommended that the Forensic Recovery of Evidence Device (FRED) from Digital Intelligence be used as the forensic workstation. More information can be found at http://www.digitalintelligence.com/products/fred/. Figure 2 depicts an image of the FRED. As can be seen this is not a typical workstation and has many features needed to perform a forensics investigation. In Addition to the FRED the system also needs a copy of Forensic Tool Kit (FTK) for performing the investigations and analysis. The FRED cost $6,000 and FTK for one year is $4,000.

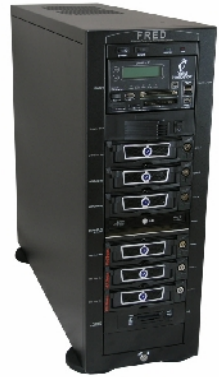

Figure 1  Forensic Recovery of Evidence Device [  HYPERLINK \l "Dig12"  **16** ] (FRED)

## 6.3   Jump Kit

The Jump Kit contains the tools and utilities needed to respond to an incident. The jump kit should be stored in a small suitcase or backpack and it should contain the following equipment:

Table 2 Jump Kit Contents

| Item | Cost | Purpose |
|------|------|---------|
| **Traveler's Choice Rome 29 in. Hardshell Spinner Suitcase** | $90 | Used to hold and transport Jump Kit Equipment |
| **Dell Latitude E6220** | $800 | Laptop to be used for Data Gathering and analysis tools |
| **Cat 6 Network cables (including crossover)** | $50 | |
| **Cisco 2960CG-8TC-L Gigabit Switch with sfp Fiber uplink** | $700 | Used to allow sniffing of the network activity. |
| **Peripheral cables—USB, Firewire, parallel, serial, console** | $100 | Used to connect to various devices and provide data transfers. |
| **BackTrack Live DVD** | $0 | Contains a clean version of Linux with many pre-installed analysis and forensic tools |
| **Notebooks, pens, pencils** | $50 | Notebooks should have numbered pages. |

| Item | Cost | Purpose |
|---|---|---|
| **Camera** | $100 | Used to photograph area and state of the computer upon arrival. |
| **1 Terabyte Hard Drive** | $120 | Used to get forensic images. |
| **UltraBlock eSATA IDE-SATA Write Blocker** | $400 | Used during the acquisition of an image to prevent writing to the original hard disk. |
| **Flashlight** | $20 | |
| **Screwdrivers, Pliers, small Wrenches** | $100 | Used to open cases and systems. May be needed for disconnecting or removing system. |
| **Call List** | | A list with the names and contact information of all Team Members, Additional Resources, and outside contacts. |

In addition to this the kit shall also have a Corporate Credit Card with $5,000 limit.

## 6.4   Recommend Software tools.

a)   Linux/Unix tools

- R-Linux
- BackTrack
- dd command
- SleuthKit
- WireShark

b)   Windows Tools

- VMWare Player
- ProDiscover
- Hex Edit
- WireShark
- steghide
- stegdetect
- Sawmill
- FINALeMAIL

## 6.5   Additional Resources

### 6.5.1 HUMAN RESOURCES.

The HR department should be involved in cases of policy violation, illegal activity, or both.

**6.5.2 FINANCE**

Will be used in the event resources, supplies, or equipment must be purchased

**6.5.3 LEGAL**

The Legal department should be contacted and involved in cases of data breach, illegal activity, and when legal questions arise.

**6.5.4 CEO**

It is imperative that the CEO be involved in major incidents; CAT 1, CAT 2, and CAT 3 from Table 2. These incidents have a severe impact on the operation and reputation of the organization as well as legal and personnel implications.

**6.5.5 PHYSICAL SECURITY**

Physical Security needs to be involved for cases of theft, unauthorized physical access, and cases where employees are to be escorted off the premises.

**6.5.6 PUBLIC RELATIONS**

Public relations must get involved when there will be a press release or breach notifications.

**6.5.7 LAW ENFORCEMENT**

In the event that there is evidence of a crime the agency should have law enforcement contacts documented. Law enforcement should only be brought in by the legal department or the CEO; unless the crime involves those individuals.

# 7 Initial Reporting.

When reporting an incident the following information is required. Sample form in Appendix
- Point of contact information including name, telephone, and email address
- Incident Category Type (e.g., CAT 1, CAT 2, etc., see table)
- Incident date and time
- Source IP, port, and protocol (If Known)
- Destination IP, port, and protocol (If Known)
- Location of the system(s) involved in the incident
- Method used to identify the incident (e.g., IDS, audit log analysis, system administrator, end-user complaint)
- Description of the Incident.

# 8 Classifying and Prioritizing Incidents

i. The US-CERT lists that reporting categories for US Agencies along with reporting time requirements [**17**]. A category 1 incident carries the highest priority while a category 6 carries the lowest priority. It is useful to use these categories as a guide to assigning urgency remembering this is just a guideline. These categories are listed and described in table 3. It should be noted that for purposes of this organization CAT 3 and CAT 4 have been exchanged.

Table 3 Incident Categories

| Category | Name | Description |
|---|---|---|
| CAT 1 | Unauthorized Access | In this category an individual gains logical or physical access without permission to a federal agency network, system, application, data, or other resource |
| CAT 2 | Denial of Service (DoS) | An attack that *successfully* prevents or impairs the normal authorized functionality of networks, systems or applications by exhausting resources. This activity includes being the victim or participating in the DoS |
| CAT 3 | Improper Usage | A person violates acceptable computing use policies. |
| CAT 4 | Malicious Code | *Successful* installation of malicious software (e.g., virus, worm, Trojan horse, or other code-based malicious entity) that infects an operating system or application. |
| CAT 5 | Scans/Probes/ Attempted Access | This category includes any activity that seeks to access or identify a computer, open ports, protocols, service, or any combination for later exploit. This activity does not directly result in a compromise or denial of service |
| CAT 6 | Investigation | *Unconfirmed* incidents that are potentially malicious or anomalous activity deemed by the reporting entity to warrant further review. |

Table 4 prioritizes the various systems and identities the risk. Criticality is a measure of the importance to the operation of the hospital. Risk is a measure of how susceptible the system is.

Table 4 – System Criticality

| System | Function | Criticality | Risk |
|---|---|---|---|
| Email | Used for internal communications between staff. External communications to vendors and partners. PPI and PHI are not authorized using email. The email server is located in the DMZ and is running as a bastion host. The only services ports open are port 25 | low | high |
| Patient Care | The patient care system includes patient check in, as well as all data pertaining to treatments. This includes scheduled services (i.e. X-Ray, Physical therapy, Lab, etc), prescribed medications, special instructions. Medications | high | low |
| Finance and Accounting | The Finance and accounting systems interface with numerous insurance companies as well as the credit card processing facility | medium | medium |

| System | Function | Criticality | Risk |
|--------|----------|-------------|------|
| **Human Resources** | Human Resources systems contain the information on all employees including their PPI, training, qualifications, work schedules. | low | low |
| **Public Web Server** | The Public Web server only provides basic information about the hospital and its services. There is no sensitive information on this system – the system does not process payments | low | medium |
| **Facilities** | The facilities engineering system monitors and controls the environmental systems within the hospital. It also monitors fuel levels, generator status, and levels of oxygen in the storage tanks. It is not connected to he internet, there is a dial up system for remote management. | high | low |
| **Desk Top** | Desktop systems are in use by ordinary users who may not recognize the risk or attempted attack. Typically, an attack on a desktop system will enable an attacker to gain access to other resources. | medium | high |

Using tables 3 and 4 as a guide the analysts can now score incidents. Start by assigning values as follows high=1, medium=2, low =3; CAT 1=1, CAT 2=2, etc. Criticality to the operation is of more importance than the risk or exposure. To obtain the score for the incident use the formula score = CAT + (2 x Criticality) + risk. These scores can then be used as a guide in prioritizing incidents.

# 9   Conclusion

An incident response team is vital to the operation of any organization. The organization does not necessarily need all of the capabilities for a complete response, but it should know where to turn for help when an incident occurs. Realizing that a cyber-incident is actually a cyber-emergency, one can then apply the proven techniques of emergency response and emergency management to the cyber incident.

The Incident Command System (ICS) has been in use in the Emergency services field since the 1970's . This model can be easily adapted to cyber incidents to allow a proper response to any size incident. The major emphasis must be on planning, this allows the organization to determine its abilities and limitations as well as identify additional resources that may be needed. Proper planning will allow the organization to respond rapidly, properly, and provide for the proper mitigation and analysis. The time to plan is before the incident, not during.

## Author

**Charles DeVoe** is a CERT analyst with the Center for Internet Security, MS-ISAC division.  Charles has a Masters of information Security from Capella University and a Bachelor of Science in Electrical Engineering from Union college, Schenectady NY. Current interest are information security, incident response, malware analysis, mobile device forensics, and computer systems forensics 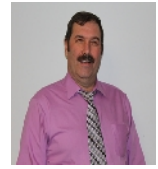

**Dr. Syed (Shawon) M. Rahman** is an assistant professor in the Department of Computer Science and Engineering at the University of Hawaii-Hilo and an adjunct faculty of School of Business and Information Technology at the Capella University.  Dr. Rahman's research interests include software engineerin g education, data visualization, information assurance and security, web accessibility, software testing and quality assurance. He has published more than 85 peer-reviewed papers. He is a member of many professional organizations including ACM, ASEE, ASQ, IEEE, and UPE. 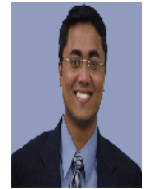